\documentstyle[prl,aps,graphicx,multicol]{revtex}
\begin{document}
\title{\bf Exact Thermodynamics of Disordered Impurities in Quantum 
Spin 
Chains}
\vspace{1.0em}
\author{A. Kl\"umper$^a$ and A. A. Zvyagin$^{a,b}$}
\address{$^a$Institut f\"ur Theoretische Physik, Universit\"at zu 
K\"oln, 
Z\"ulpicher Str. 77, D-50937 K\"oln, Germany}
\address{$^b$B. I. Verkin Institute for Low Temperature Physics and 
Engineering  \\ of the National Academy 
of Sciences of Ukraine, 47 Lenin Ave., Kharkov 310164, Ukraine}
\maketitle
\begin{abstract}
Exact results for the thermodynamic properties of ensembles of 
magnetic 
impurities with randomly distributed host-impurity couplings in the 
quantum 
antiferromagnetic Heisenberg model are presented. 
Exact calculations are done for arbitrary values of 
temperature and external magnetic field. 
We have 
shown that for strong disorder the quenching of the impurity moments 
is 
absent. For weak disorder the screening persists, but with the 
critical 
non-Fermi-liquid behaviors of the magnetic susceptibility and 
specific heat. 
A comparison with the disordered Kondo effect experiments in dirty 
metallic 
alloys is performed. 
\end{abstract}
\centerline{PACS numbers 75.10.Nr, 71.10.Hf, 75.20.Hr, 75.30.Hx}

\begin{multicols}{2}     
\narrowtext

In the last few years the interest in non-Fermi-liquid (NFL) behavior 
of 
magnetic metals and metallic alloys has grown considerably. A large 
class of 
conducting non-magnetic materials does not behave as usual Fermi 
liquids 
(FL) at low temperatures. One of the best known examples of such a 
behavior 
is the Kondo effect for multi-($n$) channel electron systems: For an 
impurity spin less than ${n\over2}$ a NFL critical behavior 
results~\cite{obz}. The critical behavior of a single magnetic impurity can 
also 
be connected with a quadrupolar Kondo effect or non-magnetic two-
channel Kondo effect~\cite{obz}. However, for most of the dirty 
metals and alloys, in which the NFL behavior was observed, see 
e.g.~the recent review \cite{vL} and Refs.~\cite{exp,exp2'},
the magnetic susceptibility ($\chi$) and low temperature specific 
heat ($C$) 
usually show logarithmic or very weak power law behavior with 
temperature ($T$). The resistivity linearly decreases with 
temperature 
showing a large residual resistivity. The last property together with 
the 
alloy nature of the compounds suggests that the disorder (a random 
distribution of localized $f$-electrons or a random coupling to the 
conducting electron host) may play the main role in the low 
temperature 
NFL character of such systems. The idea of (non-screened) local 
moments 
existing in disordered metallic systems has been already pointed out 
recently~\cite{BF,DKK,MDK}. It was proposed that near the metal-insulator 
transitions 
(or for the sufficiently alloyed systems far from the quantum 
critical point) 
disordered correlated metals contain localized moments. The random 
distribution of their magnetic characteristics (i.e.~their Kondo 
temperatures, $T_K$, which are characteristic energy scales for the 
crossover 
between the screened, or strong-coupling regime, and the weak 
coupling 
behavior) may be connected either with the randomness of exchange 
couplings 
of itinerant electrons with the local moments~\cite{DKK}, or with 
the 
randomness of the densities of conduction electron states~\cite{BF}. 
In 
Ref.~\cite{exp2'} the results of the measurements of the magnetic 
susceptibility, NMR Knight shift and low temperature specific heat 
have been 
reported. To explain the features of the behavior it was necessary to 
assume 
weak disorder, with a Gaussian distribution of the Kondo 
temperatures. 
However the model, which was used for the explanation of the 
experiment, was 
oversimplified~\cite{exp2'}: The magnetization of a single magnetic 
moment 
was approximated by the Brillouin function $B(ah/T+bT_K)$, where $h$ 
is the 
external magnetic field, and $a,b$ are constants. It was mentioned 
in~\cite{exp2'} that the data for the specific heat and Knight shift 
did not 
agree with the ones predicted by this simple theory, especially for 
nonzero values of the magnetic field. The disagreement could be 
caused either 
by an inadequate representation of the Kondo magnetization by the 
single-impurity theory, or by the simple replacement $T \to T + bT_K$ 
in the 
Brillouin function, or because of the not perfectly symmetric 
Gaussian 
distribution of the impurity couplings. The inhomogeneous magnetic 
susceptibility was confirmed very recently in~\cite{exp5} by muon 
spin 
rotation experiments. The role of the long-range (RKKY) coupling 
between the 
local moments was taken into account recently in~\cite{CNCJ} 
(Griffiths 
phase theory). The latter gives a qualitatively similar behavior as 
for 
models with non-interacting local moments~\cite{exp5}. 

It is known that the behavior of a single magnetic impurity in a 
one-dimensional (1D) antiferromagnetic (AF) Heisenberg spin 
$S={1\over2}$ 
chain as well as the behavior of a single Kondo impurity in a 3D free 
electron host are described by similar Bethe ansatz 
theories~\cite{obz,AJ,FZ}, e.g.~the magnetization and the 
low-temperature 
magnetic 
specific heat of the impurity for both models coincide. The 
spin-${1\over2}$ 
Heisenberg model is the seminal model for correlated many-body 
systems. Most 
of its static properties are exactly known. The spin-${1\over2}$ 
magnetic 
impurity manifests the total Kondo screening with FL-like low 
temperature 
behavior of the magnetic susceptibility and specific heat. In other 
words, 
the moment of the impurity is quenched by the host spins, like the 
one by the 
spin degrees of freedom of conduction electrons for the Kondo effect. 
On the 
other hand, for the integrable lattice models one can incorporate a 
finite 
concentration of magnetic impurities~\cite{many} without destroying 
the 
exact solvability (it is impossible in the free electron 
host~\cite{obz}, 
where a single magnetic impurity can only be embedded). Hence, for 
the random 
distribution of magnetic impurities we can suppose that low 
dimensionality is 
not principal for the Kondo screening. The absence of the magnetic 
ordering 
in the NFL Kondo systems~\cite{vL} also confirms this assumption and 
leads 
to the goal of our present investigation: To find exactly the 
thermodynamics 
of the disordered ensemble of spin-${1\over2}$ magnetic impurities in 
the 
Heisenberg chain (with various {\em random distributions of the 
impurity-host 
couplings}) for {\em arbitrary} values of the external magnetic field 
and temperature. It is the first study in which the thermodynamic 
characteristics of a {\em disordered interacting} many-body system 
are 
calculated {\em exactly} without any approximations. In this Letter 
we show 
that: (i) For several kinds of strong disorder of the impurity-host 
couplings the (Kondo) screening is absent; (ii) for weaker disorder 
the 
quenching persists, but with a NFL behavior of the magnetic 
characteristics. 
We performed a comparison of our exact results with previous 
approximate 
ones and with experimental data of NFL alloys. 

We investigate the thermodynamics of the quantum spin-${1\over2}$ AF 
chain 
with spin-${1\over2}$ impurities. The Hamiltonian of the system has 
the form 
$H = \sum_j H_{j,j+1} + H_{imp}$, where the host part is $H_{j,j+1} = 
{\vec S}_j{\vec S}_{j+1}$ (the host exchange constant $J$ is equated 
to 
unity). The impurities' part of the Hamiltonian has the standard form 
for the 
exactly solvable lattice Hamiltonians. Suppose we have an impurity 
distribution in which impurities are not nearest neighbors, then 
for 
the impurity, say situated between sites $m$ and $m+1$ of the host we 
obtain~\cite{FZ,cor,Eck}
\begin{eqnarray}
H_{imp} &=& J_{imp}\bigl( H_{m,imp} + H_{imp,m+1} -H_{m,m+1} 
\nonumber \\
&&+ i\theta [H_{m,imp},H_{imp,m+1}]\bigr) \ , 
\label{Himp}
\end{eqnarray}
where $J_{imp} = (\theta^2 +1)^{-1}$, and $[. ,. ]$ denotes the 
commutator. 
The coupling of the impurity to the host ($J_{imp}$) is determined by 
the 
constant $\theta$. It was shown in~\cite{FZ,cor} that precisely this 
constant 
determines the effective Kondo temperature of the impurity via $T_K 
\propto 
\exp (-\pi|\theta|)$: For energies higher than this crossover Kondo 
scale 
one has the asymptotically free impurity spin ${1\over2}$, while for 
the 
lower energies the impurity spin is screened, and the usual FL-like 
behavior 
persists. In other words, $\theta$ measures the shift of the Kondo 
resonance 
of the impurity level with the host spin excitations, similar to the 
standard 
picture of the Kondo effect in the electron host. We can 
independently 
incorporate any number of such impurities into the host chain, each 
of them 
will be characterized by its own $\theta_j$. Hence we obtain an 
ensemble of 
the spin-${1\over2}$ impurities with their own Kondo temperatures. 
The 
lattice Hamiltonian Eq.~(\ref{Himp}) has additional terms, which 
renormalize 
the coupling between the neighboring sites of the host, and three-
spin 
terms. However it was shown in~\cite{cor} that in the long-wave 
limit such a 
lattice form of the impurity Hamiltonian yields the well-known form 
of the 
contact impurity-host interaction similar to the one of the usual 
Kondo 
problem~\cite{obz}. The contact impurity coupling in this 
(conformal) limit 
is also determined by the same constant $\theta$. 

\begin{figure}
\begin{center}
\includegraphics[width=0.3\textwidth]{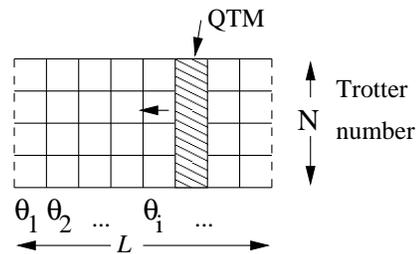}
\end{center}
\caption{Illustration of the geometry underlying the classical model 
with four-spin interaction around faces and alternating coupling 
parameters 
from column to column.}
\label{qtm}
\end{figure}

We managed to map our quantum Hamiltonian at finite temperature 
to a classical system in 2D by means of a Trotter-Suzuki 
decomposition~\cite{SuzIn}. The geometry of this classical 
system is a square lattice 
with width $L$ (=length of the quantum chain) and height $N$ 
(=Trotter 
number). 
The interactions on the lattice are four-spin interactions around 
faces with coupling parameters depending on $(NT)^{-1}$ and the 
interaction parameter $\theta_i$ where $i$ is the number of the 
column to 
which the considered face of the lattice belongs to. Note that the 
interactions are homogeneous in each column, but vary from column to 
column. We study this system in the limit $N,L \to \infty$ using an 
approach which is based on a transfer matrix describing transfer in 
horizontal direction. The corresponding column-to-column transfer 
matrices are referred to as quantum transfer matrices (QTM).
See Fig.\ \ref{qtm} for an illustration of 
the model.
In general all QTMs corresponding to the $L$ many columns are 
different. 
However, all these operators can be proven to commute pairwise. 
Therefore, 
the free energy per lattice site of our system can be calculated from 
just the largest eigenvalues of the quantum transfer matrices 
(corresponding to just one eigenstate). For a discussion 
of the homogeneous case see~\cite{K,K98} and references therein. 
For our system we find the following set of non-linear integral 
equations 
for ``energy density'' functions of spinons
$a(x)$, ${\bar a}(x)$, $A = 1 + a$ and ${\bar A} = 1 + {\bar a}$ ($x$ 
is the 
spectral parameter): 
\begin{eqnarray}
&&\int \bigl[ k(x-y)\ln A(y)- k(x-y-i\pi +i\epsilon ) \ln {\bar A}(y) 
\bigr] dy 
\nonumber \\ 
&&= \ln a(x) - {h \over 2T} + {\pi \over T\cosh x} \ ,    
\label{int}
\end{eqnarray}
with kernel function
$k(x) = {1\over 2\pi} \int d \omega {e^{-\pi |\omega| +i 2x\omega } 
/
\cosh \pi \omega }$.
The corresponding equation for ${\bar a}(x)$ is obtained from 
Eq.~(\ref{int}) 
by exchanging $i \to -i$, $h \to -h$ and $a,A \leftrightarrow {\bar 
a},{\bar 
A}$. The free energy per site $f$ is given by 
\begin{equation}
f(x) = e_0(x) - {T\over 2\pi}\int {\ln [A(y){\bar A}(y)]dy\over 
\cosh (x-y)} \ , 
\label{f}
\end{equation}
where $e_0$ is the groundstate energy. The free energy of the total 
chain 
with impurities is $F = \sum_j f(\frac{\pi}{2}\theta_j)$, where the 
sum is 
taken over all the sites (for sites without impurities we get 
$f(0)$). 
These equations are easily solved numerically for arbitrary magnetic 
field 
values and temperatures. The random distribution of the values 
$\theta_j$ 
(or of the Kondo temperatures for the impurities) can be described by 
a 
distribution function $P(\theta_j)$. The details of calculations as 
well as 
the generalization of our model for the magnetically anisotropic case 
(important for systems with strong spin-orbital 
couplings~\cite{CNCJ}) 
will be reported later.\ It is worthwhile to emphasize 
here the 
simplicity of the derived equations: For each impurity there is only 
one 
parameter, the shift of the spectral parameter in the formula for the 
free 
energy per site Eq.~(\ref{f}). Then the exact solvability of the 
problem 
for any number of impurities permits to introduce the distribution of 
these shifts (or the strengths of the impurity-host couplings, 
i.e.~the 
local Kondo temperatures). One has only two (non-linear) integral 
equations, 
Eqs.~(\ref{int}), to solve, and the answer can be obtained for arbitrary 
temperature (in principle down to $T = 10^{-24}J$)
and magnetic field ranges.

\begin{figure}
\begin{center}
\includegraphics[width=0.4\textwidth]{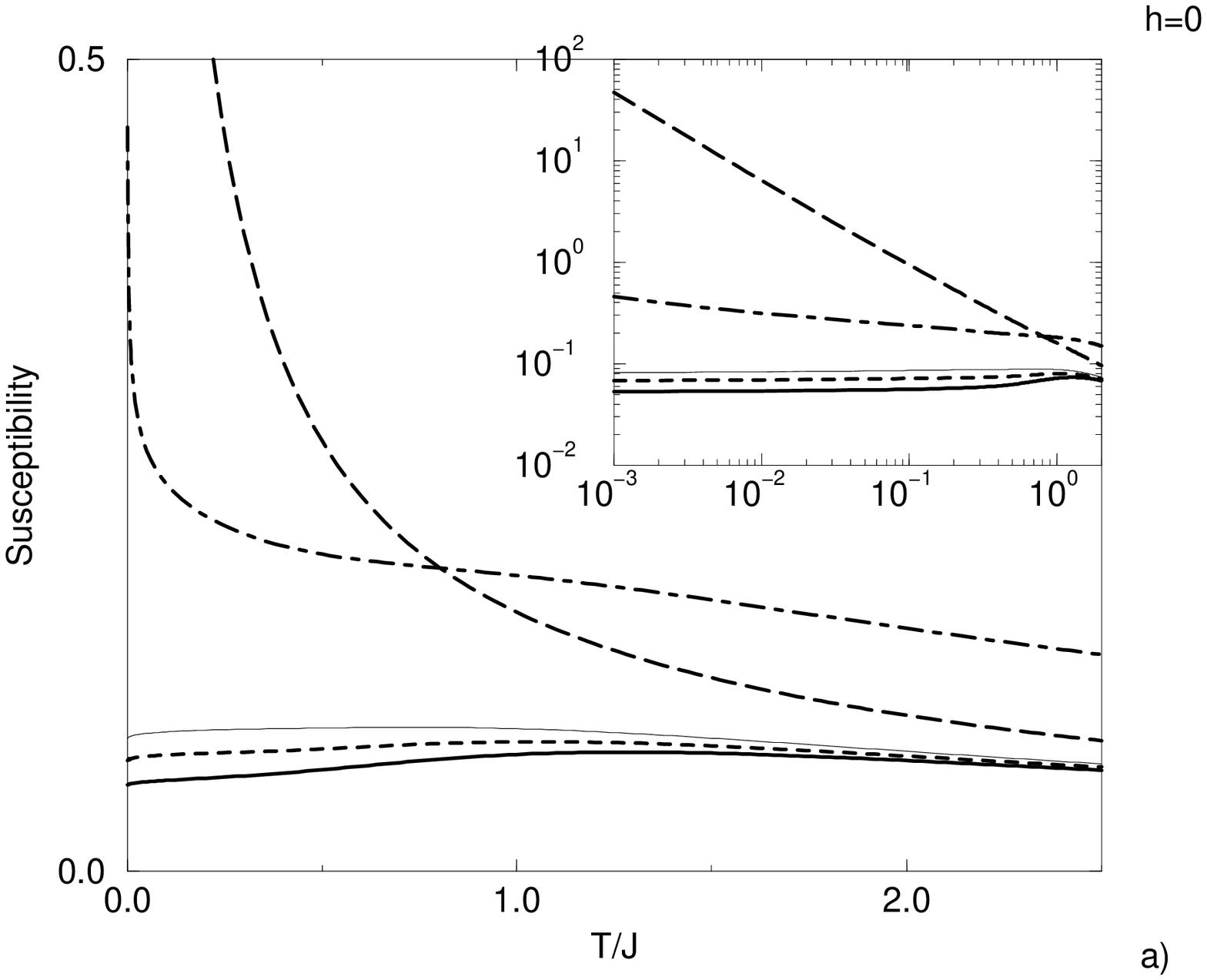}
\includegraphics[width=0.4\textwidth]{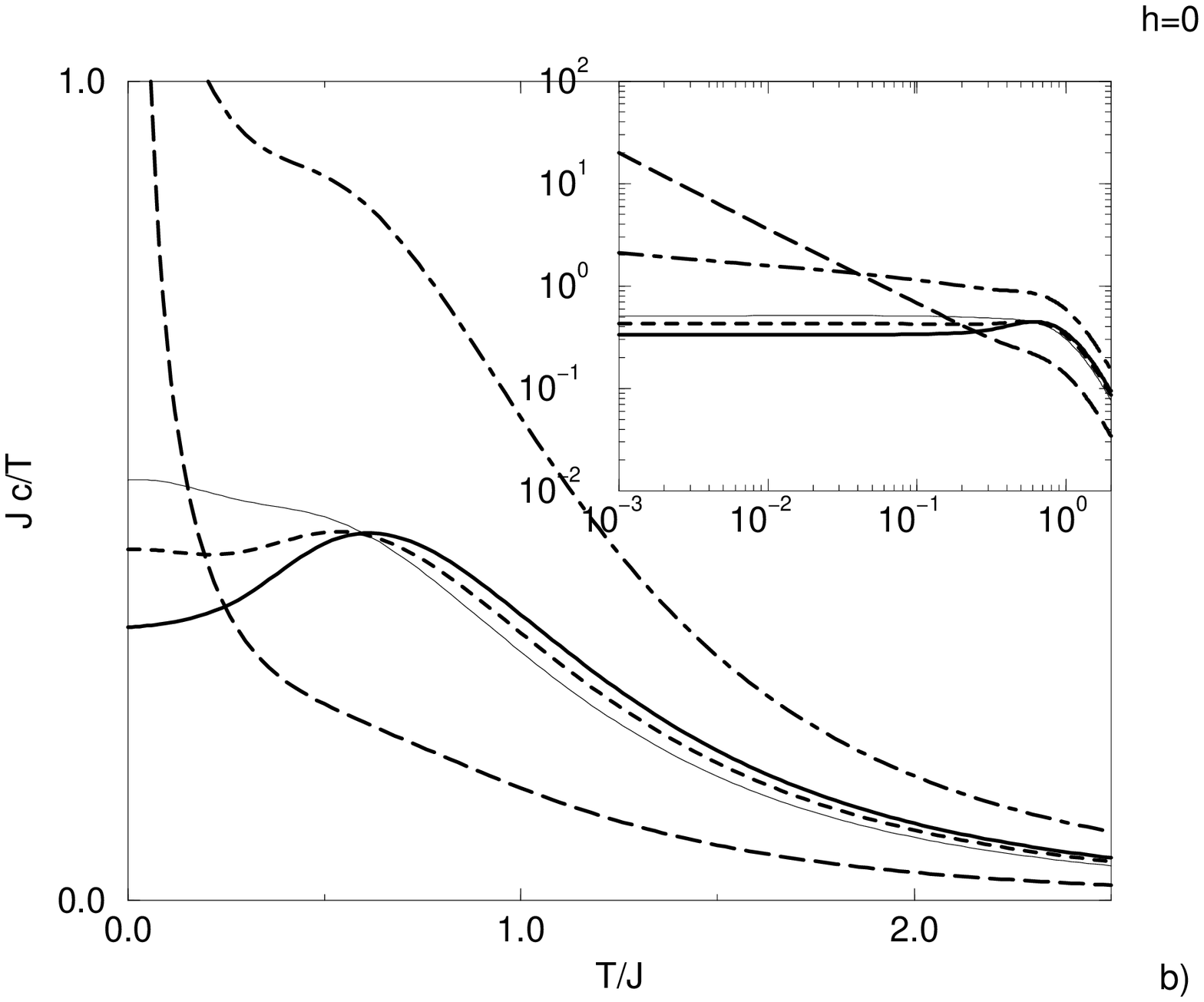}
\end{center}
\caption{The temperature dependence of (a) $\chi$ and (b) $\gamma$ at 
zero 
magnetic field for: The homogeneous Heisenberg chain (solid line); a 
single 
impurity (thin line); Gaussian distribution of the parameter 
$\theta$ 
(dashed line); Lorentzian distribution (long-dashed line); a 
log-normal 
distribution (dashed-dotted line). The insets show log-log plots of 
the 
data.} 
\label{fig1}
\end{figure}

In Fig.\ \ref{fig1} the results for (a) the magnetic 
susceptibility, and (b) the linear-temperature coefficient of the 
specific heat $\gamma = C/T$ at zero field for a homogeneous 
Heisenberg 
chain, a single Kondo impurity, a Gaussian distribution of the 
host-impurity couplings, a Lorentzian distribution, and a so-called 
logarithmically normal~\cite{AP} distribution, which is 
characteristic 
for strong disorder, e.g.~close to a critical point~\cite{DKK}) are 
plotted as functions of the temperature. One can observe a clear 
qualitative difference between strong disorder (Lorentz and 
log-normal) 
and the other curves. It is clear that the divergent value of the 
magnetic susceptibility at low $T$ for the strong disorder of the 
impurity-host couplings is connected with the fact that for most of 
the 
impurities their Kondo temperatures are lower than the temperature of 
the system. Therefore these impurities give rise to a Curie-like 
behavior 
of the susceptibility. This divergence disappears upon applying a 
finite 
external field which restors most of the FL-like behavior.

\begin{figure}
\begin{center}
\includegraphics[width=0.4\textwidth]{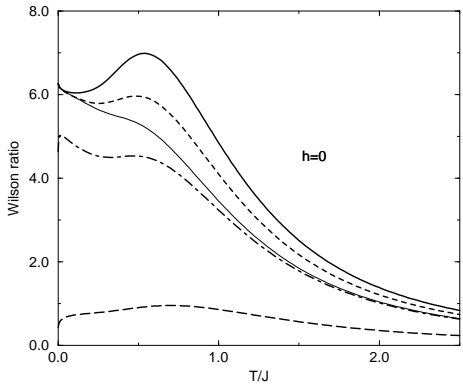}
\end{center}
\caption{The temperature dependence of the Wilson ratio~[1]
at zero magnetic 
field. Depiction of distributions by lines as in Fig.\ \ref{fig1}. 
Note the 
infinite slope at $T=0$ due to logarithmic corrections.} 
\label{fig3}
\end{figure}

Our results for the low-temperature thermodynamics are close to the 
ones of 
the perturbative calculations of random AF spin-${1\over2}$ 
chains~\cite{DM,HJ}: $\gamma$ and $\chi$ have weak power-law or logarithmic 
singularities. Our low temperature results confirm the very weak 
dependence 
of the critical exponents on the temperature. For the log-normal 
distribution we find critical exponents of 0.134 and 0.131 for 
$\chi(T)$ 
and $\gamma(T)$. In the case of the Lorentzian distribution these 
exponents 
are 0.846 and 0.730. The rather large deviations of these exponents 
are 
due to strong logarithmic corrections in $\chi(T)$ at low 
temperatures. 
For an illustration of this effect see also Fig.\ \ref{fig3} showing 
the 
Wilson ratios $\gamma /\chi$ which are non-universal, i.e.~with NFL 
behavior and show infinite slope at $T=0$. 
We like to note that 
former 
calculations were valid at most for low temperatures, while our 
approach 
is applicable for {\em any} temperature and magnetic field scales. 
Furthermore, it is also known that the approximate 
results~\cite{DM,HJ} 
incorrectly give zero or infinite susceptibility at $T \to 0$ 
irrespective of the distribution, while its true value is finite 
for e.g.~the homogeneous chain and the single 
impurity~\cite{FZ,K98}. Our 
scheme, of course, perfectly describes the correct behavior. 
We also performed a comparison of our results with the data of 
Ref.~\cite{exp2'}, with qualitatively similar results. 
Note that any distribution $P(\lambda)$ of impurity couplings 
$\lambda (\propto 1/\theta$ in our parametrization)
with finite $P(0)$ 
corresponds to a distribution in $\theta$ with Lorentzian tail. Hence 
the agreement of our data with \cite{exp2',MDK}: infinite $\chi$ 
and $\gamma$ for $T \to 0$.  
More clearly the 
comparision with experiments can be seen from Refs.~\cite{exp6,Bul}. 
By 
changing the concentration of impurities one can go from weak to 
strong 
disorder with quantative agreement with wide and narrow Gaussian and 
log-normal distributions~\cite{exp6}. Critical exponents very close
to those of our Lorentzian distribution were observed in~\cite{Bul}.

Generalizing model (1) by keeping exact solvability, it is possible 
to include (random) short- and 
long-range antiferromagnetic interactions of the special forms 
between the impurities themselves, see e.g.~\cite{many,int}. 
However, these interactions do not affect the 
behavior of our disordered correlated spin system qualitatively, 
compared to the case without direct interaction between impurities. 
Here, a probability distribution $P(\theta)$
with asymptotics $|\theta|^{-\alpha}$ for large $\theta$ leads to
a divergence $\chi(T)\propto T^{-1/\alpha}$ which is weaker
than that observed above. It was 
suggested in~\cite{CNCJ} that the inclusion of other kinds of 
impurity-impurity couplings (of RKKY form, which violate the exact 
integrabilty) also do not change the behavior qualitatively. 
However ferromagnetic impurity-impurity couplings can change the 
situation drastically, e.g.~providing infinite $\chi (T\to 0)$ 
even for weak disorder. 
The generalization of our results to correlated electron systems
with random impurities will be reported elsewhere.

To conclude, we have constructed exactly the thermodynamics of the 
Heisenberg antiferromagnetic spin-${1\over2}$ chain with embedded 
disordered impurities. The results are of high (numerical) accuracy 
and 
valid for arbitrary ranges of magnetic field and temperature. For 
strong 
disorder of the impurity-host couplings the local moments are 
non-quenched. 
For weak disorder of the host-impurity couplings, on the other hand, 
spin 
excitations of the host screen the impurities, but with the 
non-Fermi-liquid 
behaviors of the thermodynamic characteristics. The comparison of our 
theory 
with the data of a perturbative analysis and with those of magnetic 
experiments on disordered non-Fermi-liquids in rare-earth alloys 
shows 
qualitative agreement. 

Support by {\it Deutsche Forschungsgemeinschaft} and
{\it Sonderforschungsbereich} 341, K\"oln-Aachen-J\"ulich is acknowledged. 
A. A. Z. thanks the Institut f\"ur Theoretische Physik, Universit\"at 
zu 
K\"oln for its kind hospitality.

\end{multicols}
\end{document}